\documentclass[aps,pra,twocolumn,superscriptaddress]{revtex4-1}
\usepackage{graphicx}
\usepackage{amsmath,amsfonts}

\begin{document}


\title{New concepts of inertial measurements with multi-species atom interferometry}



\author{Alexis Bonnin}
\affiliation{ONERA-The French Aerospace Lab, F-91123 Palaiseau Cedex, France}
\author{Cl\'ement Diboune}
\affiliation{ONERA-The French Aerospace Lab, F-91123 Palaiseau Cedex, France}
\author{Nassim Zahzam}
\affiliation{ONERA-The French Aerospace Lab, F-91123 Palaiseau Cedex, France}
\author{Yannick Bidel}
\affiliation{ONERA-The French Aerospace Lab, F-91123 Palaiseau Cedex, France}
\author{Malo Cadoret}
\affiliation{ONERA-The French Aerospace Lab, F-91123 Palaiseau Cedex, France}
\affiliation{LCM, CNAM, 61 rue du Landy, 93210 La Plaine Saint-Denis, France}
\author{Alexandre Bresson}
\affiliation{ONERA-The French Aerospace Lab, F-91123 Palaiseau Cedex, France}


\date{\today}

\begin{abstract}
In the field of cold atom inertial sensors, we present and analyze innovative configurations for improving their measurement range and sensitivity, especially attracting for onboard applications. These configurations rely on multi-species atom interferometry, involving the simultaneous manipulation of different atomic species in a unique instrument to deduce inertial measurements. Using a dual-species atom accelerometer manipulating simultaneously both isotopes of rubidium, we 
report a preliminary experimental realization of original concepts involving the implementation of two atom interferometers first with different interrogation times and secondly in phase quadrature.
These results open the door to a new generation of atomic sensors relying on high performance multi-species atom interferometric measurements.
\end{abstract}

\pacs{}

\maketitle

\section{Introduction}
Cold atom interferometry has allowed in the last decades the development of extremely sensitive and accurate inertial sensors for measuring the gravity acceleration \cite{Peters2001}, Earth's gravity gradient \cite{McGuirk2002} or rotations \cite{Gustavson1997}. They appear very promising for a wide range of applications such as inertial navigation \cite{Jekeli2005}, geodesy \cite{Carraz2014}, natural resource exploration \cite{Nabighian2005}, or fundamental physics \cite{Bouchendira2011,Fixler2007,Dimopoulos2007,Dimopoulos2008,Muller2008,Rosi2015}. To address most of these applications, an important research effort is made to make these cold atom instruments more adapted to withstand operational constraints \cite{VanZoest2010,Geiger2011,Aosense,Muquans}. In addition to improve these instruments in terms of compactness and robustness, it is of major concern to extend their dynamic measurement range. When shot-to-shot acceleration variations are small compared to the dynamic measurement range, defined here as one half of the interferometer fringe spacing,  cold atom gravimeters allow to retrieve the gravity acceleration $g$ with a resolution of $\approx 10^{-9}g$ at a repetition frequency of few Hz. This dynamic measurement range lies in the range 0.2 - 3 $\mu g$ for state-of-the-art cold atom gravimeters \cite{Hu2013,Gillot2014,Freier2015}, much less than the acceleration level of roughly 50 $\mu g$ that could be encountered in a typical laboratory environment, 50 $mg$ that could be encountered in a plane \cite{Geiger2011} and 0.3 $g$ that could be encountered in a ship \cite{Bidel2017}. To allow the operation of these atomic sensors in a laboratory environment, they are typically used on a passive \cite{LeGouët2008,Bidel2013} or active vibration isolation platform \cite{Hensley1999,Zhou2012,Hauth2013} or set directly on the ground by combining the atomic signal with that of a conventional accelerometer \cite{Merlet2009,Lautier2014,Geiger2011,Bidel2017}. To our knowledge, only few onboard demonstrations of inertial measurements have been reported whether in an elevator \cite{Bidel2013}, on a plane \cite{Geiger2011}, or on a boat \cite{Bidel2017}. For the elevator demonstration, the interrogation time of the atom interferometer is reduced significantly to increase the fringe spacing, and therefore the dynamic measurement range, but at the cost of a drastic loss of the sensitivity measurement. For the plane and boat demonstrations, the atom interferometer is coupled to a conventional accelerometer which helps to determine, for each measurement cycle, the fringe index on which the atom interferometer is operating. In this last configuration, the use of a conventional accelerometer often limits the performance of the instrument either because of the lower performance intrinsic noise of the conventional accelerometer, its non-linearity, the lack of knowledge of its transfer function or because of a mismatch between the localization of the conventional accelerometer measurement point and that of the atom interferometer.

As an alternative or in complement to vibration isolation platforms and conventional accelerometers, we propose and study in this article new concepts of an atomic accelerometer dedicated to onboard applications. These concepts could be interesting in the future for increasing the dynamic measurement range of the atomic instrument and allowing the atomic interferometer to operate always in its linear regime \cite{Merlet2009}, leading to a maximum sensitivity whatever the measured acceleration. The instrument configurations that will be presented do not rely on external devices making our instrument self-independent and only relying on atom interferometry. These innovative concepts are based on simultaneous atom interferometry with different atomic species. Up to now, the few experiments dealing with multi-species atom interferometry were dedicated to test the universality of free fall \cite{Bonnin2013,Schlippert2014,Tarallo2014,Zhou2015,Duan2016,Barrett2016} and their interests to improve single species atomic inertial measurements were never discussed to our knowledge. The principle of the experiment is the following. The atomic species are trapped and cooled simultaneously in the same vacuum chamber, sharing the same laser beams and constituting a whole unique embedded atomic cloud. The atomic species are then free falling simultaneously, feeling therefore the same inertial forces, and during which they are interrogated through atom interferometry by the same laser beam. At the end of the free fall, each atomic species is detected and provide complementary acceleration measurements that allow to improve performances reached by standard single species atom interferometers. Using different atomic species in the same vacuum chamber is of great interest since each atom-light interaction for manipulating one atomic species should not disturb significantly the other atomic species. Indeed each laser line dedicated to manipulate one atomic species is seen far of resonance regarding the other atomic species.\\

\section{Presentation of the multi-species concepts}
\subsection{Introduction}
In a general case, the phase shift $\Delta\Phi$ accumulated between both paths of a Mach-Zehnder type interferometer is proportional to the acceleration of the atoms $a$ along the interferometer's laser direction of propagation \cite{Peters2001}:
\begin{equation}\label{interferometer phase}
\Delta \Phi = k_{\mathrm{eff}}\cdot a \cdot T^{2}
\end{equation}
The interferometer is characterized by three laser pulses of effective wave-number $k_{\mathrm{eff}}\approx \frac{4\pi}{\lambda}$, where $\lambda$ is the wavelength of the laser, and separated by the interrogation time $T$.\\
Experimentally, the measured quantity at the output of the interferometer is the transition probability $P$ to be in a particular atomic state which is a sinusoidal function of the interferometric phase:
\begin{equation}\label{transition probability}
P = P_{0}-\frac{A}{2}\cos\left(\Delta\Phi + \Delta\Phi_{\mathrm{op}}\right)
\end{equation}
where $P_{0}$ and $A$ are respectively the offset and the amplitude of the fringes. $\Delta\Phi_{\mathrm{op}}$ is a very well known and controlled phase shift added by the operator. For instance, in an atom gravimeter, a frequency chirp $\alpha$ is added to the interferometer's laser frequency to compensate the Doppler shift induced by gravity $g$ such as $\Delta\Phi_{\mathrm{op}}=2\pi\alpha T^{2}=-k_{\mathrm{eff}}\cdot g\cdot T^{2}$.

The sinusoidal response of the atom interferometer is responsible of two important drawbacks when the phase shift $\Delta\Phi$ is extracted.

\begin{figure}
\includegraphics[width=8.5cm]{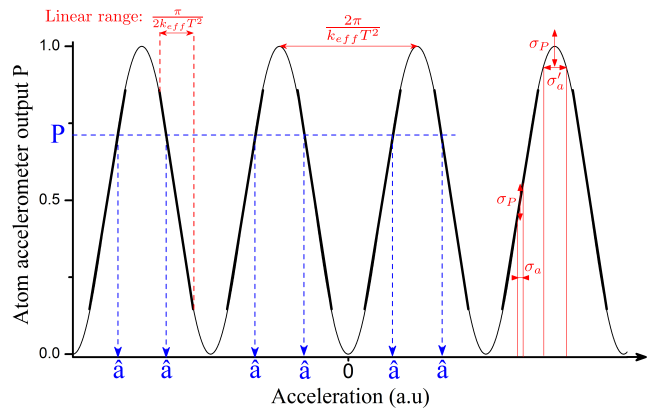}
\caption{Representation of a typical output signal from an atom accelerometer. The measured signal $P$ corresponds to a large number of possible acceleration values $a$. The interferometer operates at its maximum sensitivity in a restricted linear range, away from the fringe extrema where $\left|dP/d\Delta\Phi\right| = 0$. Considering an intrinsic noise of the interferometer $\sigma_P$, the associated acceleration noise when operating on the fringe extrema $\sigma^{\prime}_{a}$ is larger than the associated acceleration noise when operating in the linear regime $\sigma_a$.\label{figure1}}
\end{figure}

First, when shot-to-shot variation of the acceleration signal induces phase shifts larger than $\pi$, then the measured transition probability $P$ could correspond to a large number of acceleration values (see Fig.\ref{figure1}):
\begin{equation}\label{acceleration possible values}
\hat a=\frac{s}{k_{\mathrm{eff}}T^{2}}\arccos\left(2\frac{P_{0}-P}{A}\right)-\frac{\Delta\Phi_{\mathrm{op}}}{k_{\mathrm{eff}}T^{2}}+n\frac{2\pi}{k_{\mathrm{eff}}T^{2}}
\end{equation}
\indent where $n\in\mathbb{Z}$ and $s=\pm 1$ correspond to the fringe indexes of the atom interferometer at each cycle. For onboard applications, a conventional accelerometer is typically used to determine these fringe indexes at each cycle \cite{Geiger2011,Bidel2017} or could be used in a real time  phase compensation loop \cite{Lautier2014}.

Secondly, the sensitivity of the interferometer depends on the value of the total phase shift and consequently on the measured acceleration. When $\Delta\Phi+\Delta\Phi_{\mathrm{op}}=n\pi$ ($n$ being an integer), \textit{i.e} when the interferometer works at the top or at the bottom of the fringes, the sensor becomes highly insensitive because the derivative of the measured probability $P$ with respect to $\Delta\Phi$ is zero (see Fig.\ref{figure1}). In a low vibration environment, cold atom gravimeters are usually forced to operate at mid-fringe by adjusting $\Delta\Phi_{\mathrm{op}}$ at each cycle to  ensure a maximum sensitivity of the instrument \cite{Peters2001}. For vibration levels exceeding the measurement range of the atom interferometer, a solution consists in compensating the vibrations in real time thanks to a conventional accelerometer allowing the atom interferometer to stay at mid-fringe \cite{Lautier2014}.\\
In this article, we propose two configurations to mitigate these two issues. 

\subsection{First concept}
The first configuration relies on the implementation of two simultaneous atom interferometers with two different interrogation times $T_1$ and $T_2$ [see Fig.\ref{figure2}(a)]. The atom interferometer with the highest interrogation time $T_2$ presents a greater sensitivity but conversely a smaller measurement range. On the other hand, the atom interferometer with a lower interrogation time $T_{1}$ benefits from an increased measurement range, by a factor $\left(T_{2}/T_{1}\right)^{2}$, but at the expense of a degraded sensitivity, following the same scaling factor. Associating both interferometer signals allows to benefit simultaneously from a high sensitivity and a high measurement range. Note that, for a given measurement range fixed by the atom interferometer with $T_1$, this configuration allows also to benefit from a higher accuracy thanks to the interferometer with the highest interrogation time $T_2$. Indeed, the impact of a significant number of systematic effects scales inversely with the interrogation time.
\subsection{Second concept}
\begin{figure}
\includegraphics[width=8.5cm]{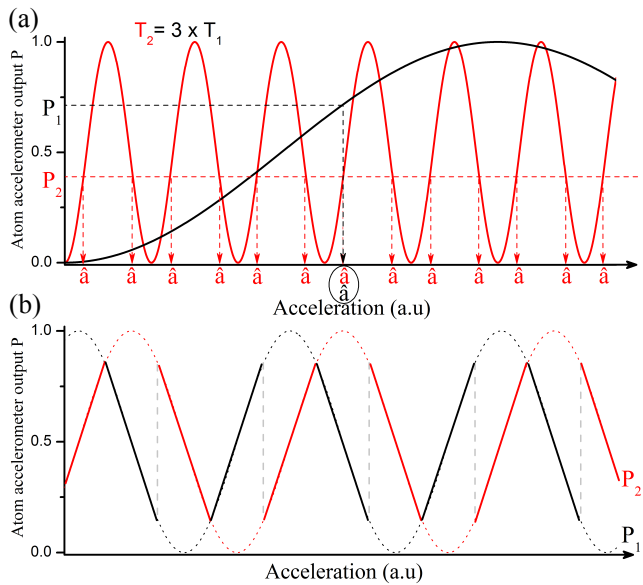}
\caption{(a) Representation of the output signals from two atom interferometers operating with two different interrogation times $T_1$ and $T_2 = 3 T_1$. The atom interferometer operating with $T_1$ benefits from a 9 times higher measurement range. (b) Representation of the output signals from two atom interferometers in phase quadrature. Whatever the acceleration, at least one atom interferometer is operating in the linear regime.}
\label{figure2}
\end{figure}

The second proposed configuration relies on the implementation of two simultaneous atom interferometers operating in phase quadrature [see Fig.\ref{figure2}(b)]. The linear ranges from both interferometers can be coupled and a fully linear response gravimeter is obtained allowing to get rid of insensitive acceleration regions where $\left|dP/d\Delta\Phi\right| \approx 0$.\\

\section{Experimental demonstrations}
\subsection{Experimental setup}
We report here preliminary experimental demonstration of these two configurations. The experimental setup is similar to the one described in Ref. \cite{Bonnin2015}. With this setup, approximatively $6\times10^{8}$, respectively $8\times10^{8}$, atoms of $^{87}$Rb, resp. $^{85}$Rb, are simultaneously loaded from a background vapor into a 3D Magneto-Optical Trap (MOT) in 250 ms. The atoms are then further cooled down in an optical molasses phase allowing the atoms to reach a temperature of $\sim$ 2 $\mu$K. The atoms are then selected in the magnetic insensitive state $|F=1,m_{F}=0\rangle$ for $^{87}$Rb and $|F=2,m_{F}=0\rangle$ for $^{85}$Rb thanks to a microwave pulse and then dropped over a distance of $\sim6$ cm. For both atomic species, a Mach-Zehnder type interferometer is implemented, consisting in three equally spaced Raman laser pulses $\pi/2-\pi-\pi/2$. Note that the interrogation times can be varied for each species to a maximum value of $T=47$ ms. The Raman laser beam is retro-reflected on a mirror which acts as the reference for the inertial measurement. Since both atom interferometers share the same Raman laser beam, the two measurement axis are therefore rigorously the same. Finally, the atomic probabilities are successively measured for each isotope by fluorescence detection. The whole sequence is performed at a repetition rate of 4 Hz. The instrument is installed on a vibration isolation platform to clearly resolve the atomic fringes and minimize the interferometer phase noise which will allow us to analyze more simply the potential of the presented concepts.
Despite an apparent complexity, the dual species experiment stays rather simple to implement and not significantly more complex than a single species instrument. All the required laser lines necessary for both isotopes manipulation are synthesized thanks to phase modulation of a single laser source \cite{Bonnin2015}. 
During the cooling stage, the carrier frequency is tuned on the $^{87}$Rb cooling transition. The $^{85}$Rb cooling line and both repumper lines are generated by three modulation frequencies injected into the phase modulator as following:
f$_{cooling}^{~85}$ = f$_{carrier}$ + 1.126 GHz, f$_{repumper}^{~87}$ = f$_{carrier}$ + 6.568 GHz,
f$_{repumper}^{~85}$ = f$_{carrier}$ + 1.126 + 2.915 GHz. Phase modulation is also used for generating the laser lines during the interferometric sequence. Both Raman pairs are generated by directly injecting the Raman difference frequencies associated to both isotopes (\textit{i.e.} 6.834 GHz for $^{87}$Rb and 3.035 GHz for $^{85}$Rb), making the carrier frequency common to both Raman pairs. The Raman pair corresponding to $^{87}$Rb is red detuned by 0.59 GHz with respect to the excited hyperfine state $F'=2$ and therefore the one corresponding to $^{85}$Rb is red detuned by 1.86 GHz with respect to $F'=3$.
Compared to the one species instrument, the dual species instrument requires only one additional microwave source fitting in a standard 3U 19 inch rack.\\
\subsection{First concept}
\begin{figure}
\includegraphics[width=8.5cm]{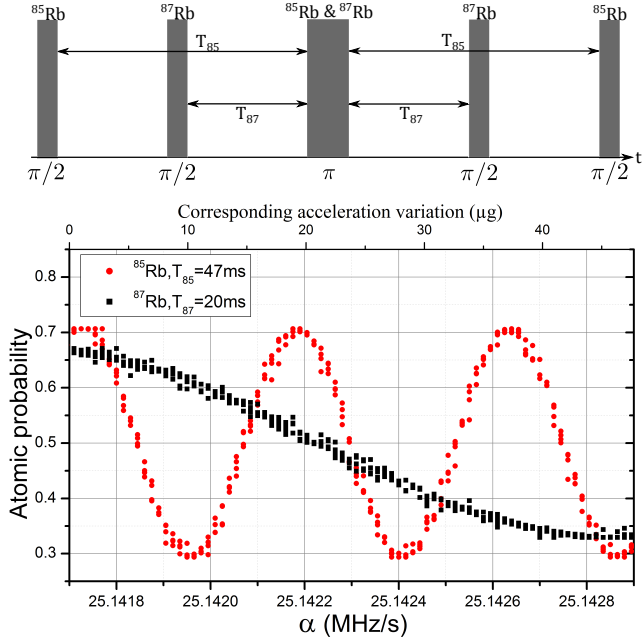}
\caption{At the top, the global interferometric time sequence for interrogating both isotopes. At the bottom, experimental interferometric signals for two simultaneous interferometers with two different interrogation times: $T_{87}=20$ ms (black squares) and $T_{85}=47$ ms (red circles). The interferometric fringes are scanned thanks to a frequency chirp $\alpha$ applied on the Raman laser. \label{figure3}}
\end{figure}

The first dual-species operating mode that we study in this article allows to increase the measurement range of a single species atom accelerometer. For the experimental realization of this operating mode, we set the interrogation time for $^{87}$Rb to $T_{87}=20$ ms and the one for $^{85}$Rb to $T_{85}=47$ ms. The central $\pi$-pulse is kept common for both isotopes (see Fig.\ref{figure3}). This feature allows to keep the global symmetry of the entire interferometric sequence in order to cancel out impact of light shifts on one atomic species due to the laser lines used to address the other isotope. Moreover, in such configuration, the effective measurement points of both interferometers are temporarily and spatially close which limits the impact of spurious rotations or gravity gradients. Nevertheless, this configuration, implementing two interferometers with different interrogation times, leads to different transfer functions for the two atomic accelerometers which will limit the efficiency of their coupling for high frequency signals. The interferometric sequence for $^{87}$Rb, respectively $^{85}$Rb, consists of three laser pulses of duration 1.3 - 8 - 1.3 $\mu$s, respectively 3.2 - 8 - 3.2 $\mu$s. The central $\pi$ pulse being the same for both isotopes, the Rabi frequencies associated to the Raman transitions are different from those of the first and last $\pi/2$ pulses, leading to small decrease of the atomic fringe contrast. The interferometric signals coming from such configuration are shown in Fig.\ref{figure3}. The fringes are scanned thanks to a frequency chirp $\alpha$ applied on the Raman laser. An increase of the measurement range by a factor $\left(T_{85}/T_{87}\right)^{2}\sim5.5$ can be clearly observed.

\begin{figure}
\includegraphics[width=8.5cm]{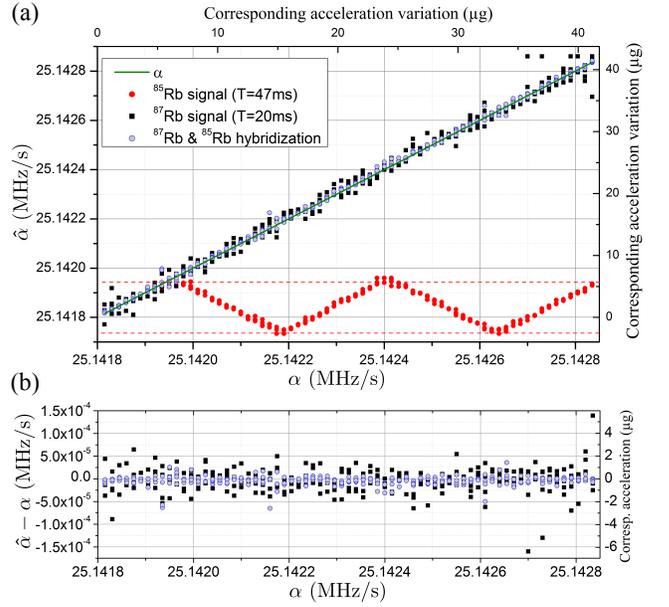}
\caption{(a) Non-ambiguous extraction of $\alpha$ from the experimental signals of Fig.\ref{figure3} where $T_{87}=20$ ms and $T_{85}=47$ ms. The frequency chirp, corresponding to a constant acceleration applied on the instrument, can be extracted from interferometric signals of $^{87}$Rb alone (black squares), of $^{85}$Rb alone (red circles) or from the hybridization of both signals (blue-grey circles). The true value of the applied frequency chirp is given in green line. The$^{85}$Rb signal alone is not sufficient to extract $\alpha$ unambiguously. (b) Residue coming from the comparison of the extracted value of $\alpha$ with its true value. Standard deviation of the residue coming from the dual species signal (blue-grey circles) is 2.6 times smaller than the one of residue coming from the single species $^{87}$Rb interferometer (black squares).
\label{figure4}}
\end{figure}

In order to highlight the interest of such configuration to allow larger acceleration variation measurements, we have extracted the value of $\alpha$ from the two experimental interferometric signals thanks to Eq.(\ref{acceleration possible values}): 
\begin{equation}\label{equationAlpha}
\hat \alpha= -\frac{k_{\mathrm{eff}} a}{2\pi} + \frac{s}{2\pi T^{2}} \arccos\left(2\frac{P_{0}-P}{A}\right) + \frac{n}{T^{2}}
\end{equation}
The extracted value of $\alpha$ corresponds to $\hat \alpha$ and $a$ can be considered constant and equal to the gravity acceleration $g$. Note that applying this frequency chirp $\alpha$ is equivalent to generating a constant acceleration of the atoms with respect to the Raman laser. The results are shown in Fig.\ref{figure4}(a). The $^{85}$Rb interferometer alone (red circles) does not allow to extract $\alpha$ in the full range because of ambiguity on the determination of the fringe indexes (see Eq. \ref{acceleration possible values}). On the other hand, the $^{87}$Rb interferometer alone (black squares), with a smaller $T$, allows to extract $\alpha$ without any ambiguity on the entire variation range. Taking into account both signals (blue-grey circles) allows to overcome the ambiguity by keeping among the possible values of $\alpha$ given by the $^{85}$Rb interferometer the value which is closest to that given by the $^{87}$Rb interferometer. 

\begin{figure}
\includegraphics[width=8.5cm]{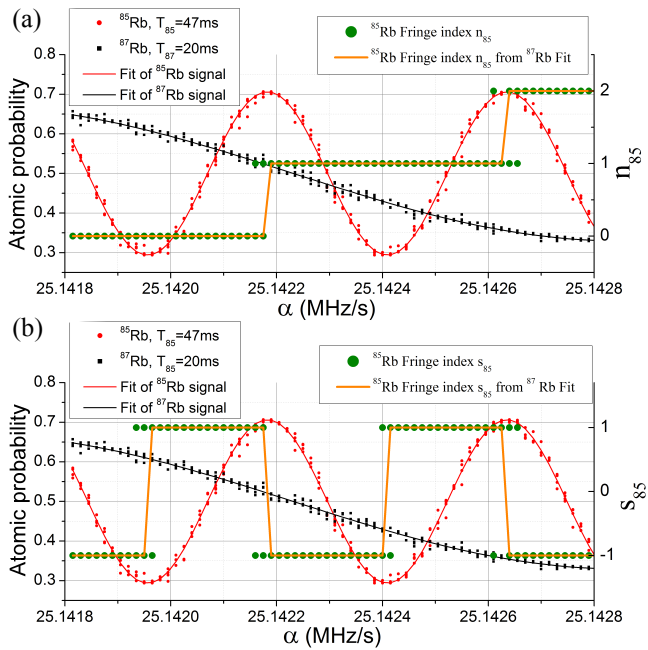}
\caption{For $T_{85}=47$ ms and $T_{85}=20$ ms, determination of the $^{85}$Rb signal (red circles) fringe indexes $n_{85}$ [green circles in plot (a)] and $s_{85}$ [green circles in plot (b)] with the help of the $^{87}$Rb signal (black squares). In orange line, the same fringe indexes have been determined using a noiseless $^{87}$Rb signal corresponding to the fitted curve of the $^{87}$Rb signal (black line).
\label{figure7}}
\end{figure}

Compared to the case of a single $^{87}$Rb interferometer with $T_{87}=20$ ms, a sensitivity gain can be seen on the residue of the extraction of $\alpha$ [see Fig.\ref{figure4}(b)]. This gain, defined as the ratio of the standard deviation of the residue associated to the $^{87}$Rb interferometer and the one of the residue associated to dual interferometer, is here equal to 2.6 and is limited by the probability noise of the $^{87}$Rb interferometer which could lead to a bad determination of the $^{85}$Rb interferometer fringe indexes, as it's shown on Fig.\ref{figure7}. In this figure, $n_{85}$ [Fig.\ref{figure7}(a)] and $s_{85}$ [Fig.\ref{figure7}(b)] correspond to the fringe indexes of the $^{85}$Rb signal used in Eq. \ref{acceleration possible values} as $n$ and $s$. The green circles give for each measurement point the values of the fringe indexes of the $^{85}$Rb signal to determine unambiguously $\alpha$. These indexes are determined by using the $^{87}$Rb signal. To evaluate the efficiency of this determination and the impact of the $^{87}$Rb probability noise on it, we have plotted on Fig.\ref{figure7} in orange line the fringe indexes values that would be determined with a noiseless $^{87}$Rb signal, using a sine function fitting the $^{87}$Rb output. 

\begin{figure}
\includegraphics[width=8.5cm]{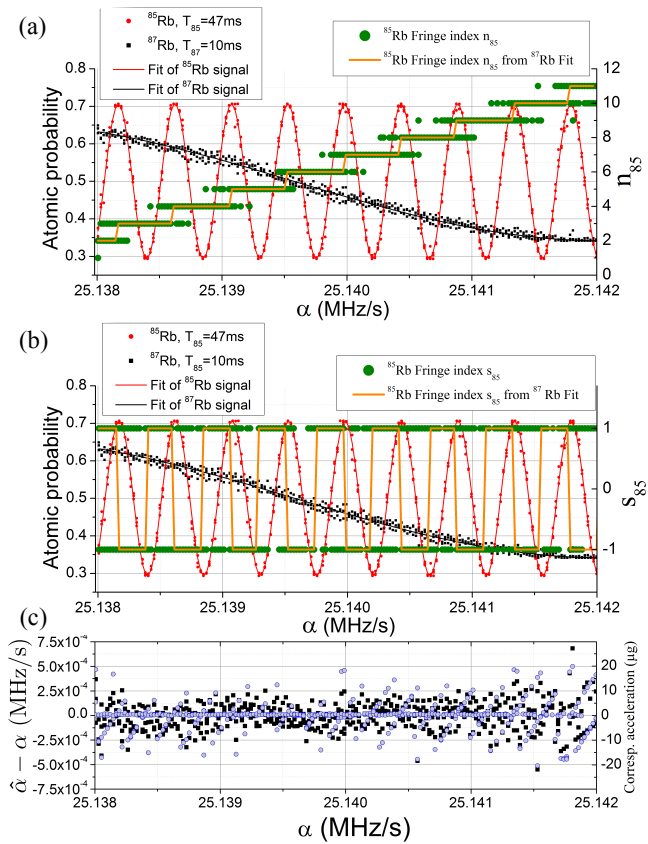}
\caption{For $T_{85}=47$ ms and $T_{85}=10$ ms, determination of the $^{85}$Rb signal (red circles) fringe indexes $n_{85}$ [green circles in (a)] and $s_{85}$ [green circles in (b)] with the help of the $^{87}$Rb signal (black squares). In orange line, the same fringe indexes have been determined using a noiseless $^{87}$Rb signal corresponding to the fitted curve of the $^{87}$Rb signal (black line). (c) The residue coming from the comparison of the extracted value of $\alpha$ with its true value. Standard deviation of the residue coming from the dual species signal (blue-grey circles) is roughly equal to the one of residue coming from the single species $^{87}$Rb interferometer (black squares).
\label{figure8}}
\end{figure}

We can see that the determination of the fringe indexes is not perfect which limits the gain in sensitivity.  In the ideal case of a noiseless $^{87}$Rb signal, the sensitivity gain reaches a factor 3.8, which is still lower than what could be expected from the ratio $(T_{85}/T_{87})^2 \sim 5.5$ coming from the $T^2$ dependance of the atom interferometer scale factor. This reduced gain results mainly from measurement points near the fringe extrema where the atom interferometer sensitivity is significantly reduced. The probability noise on the experimental data is here a strong limitation of this new concept and it prevents us from further reducing the interrogation time of the $^{87}$Rb interferometer to increase the measurement range. Indeed, the shorter the interrogation time, the worse the determination of the fringe index: when $T$ is decreased, the impact of the probability noise is increased with an acceleration noise scaling as $T^{-2}$. We can see for instance on Fig.\ref{figure8} the poor efficiency of the fringe indexes determination with $T_{85} = 47$ ms and $T_{87}=10$ ms due to the probability noise of the $^{87}$Rb signal. In this case, there is no sensitivity gain with the dual atom interferometer compared to a single species interferometer while, in the ideal situation of a noiseless $^{87}$Rb signal, the expected gain sensitivity thanks to a dual species interferometer reaches now a factor 17. Even removing measurement points near the extrema does not lead to an improvement of the sensitivity gain because the probability noise of the $^{87}$Rb signal leads now to determination errors exceeding one fringe of the $^{85}$Rb signal.\\

\subsection{Second concept}
In the more specific issue of sensitivity loss due to measurement points near fringe extrema, we have also studied a second operating mode which relies on two perfectly simultaneous interferometers working in phase quadrature. Both effective Raman wave-vectors need to be the same in order to guarantee equal scale factors and thus equal fringe spacings. For the implementation of this configuration, a differential phase $\phi_{d}=\pi/2$ is experimentally created by adding during the $\pi$-pulse a controlled phase jump on the microwave source used to generate the Raman laser pair of the $^{85}$Rb.

\begin{figure}
\includegraphics[width=8.5cm]{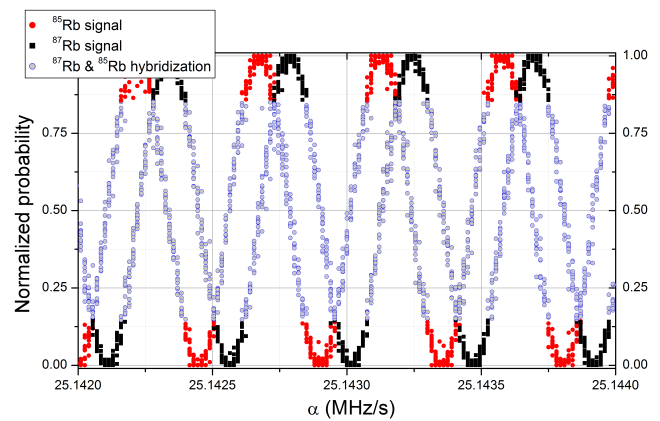}
\caption{Experimental realization of a fully linear atom interferometer. Coupling the $^{85}$Rb (red circles) and the $^{87}$Rb (black squares) interferometer outputs phase shifted by $\pi/2$, with $T_{85}=T_{87}=47$ ms and selecting the adequate signal in each phase range gives rise to a linear response sensor (blue-grey circles) with an constant optimal sensitivity.
\label{figure5}}
\end{figure}

\begin{figure}
\includegraphics[width=8.5cm]{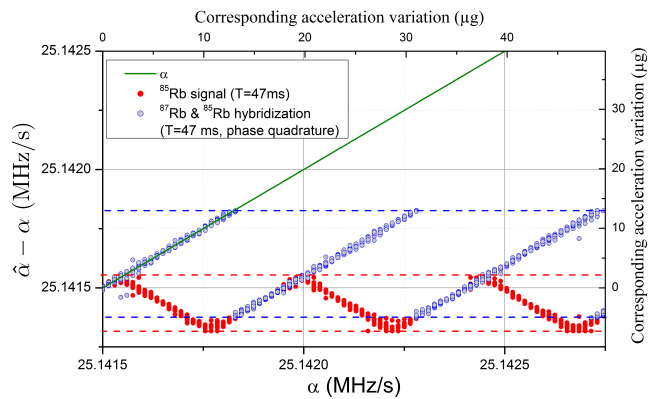}
\caption{Extraction of $\alpha$ from the experimental signals of Fig.\ref{figure5}. The frequency chirp, corresponding to a constant acceleration applied on the instrument, can be extracted from interferometric signals of $^{85}$Rb alone (red circles), of $^{87}$Rb alone (not represented on the figure but similar to the $^{85}$Rb signal with an offset due to the phase quadrature) or from the hybridization of both signals (blue-grey circles). The true value of the applied frequency chirp is given in green line. The dashed lines show the measurement range associated to the single species $^{85}$Rb interferometer (in red) and the measurement range associated to the dual species $^{85}$Rb/$^{87}$Rb interferometer in phase quadrature (in blue). For the dual species signal, only data in the linear range of each single atom interferometer have been kept, excluding therefore low sensitive data near fringe extrema.
\label{figure6}}
\end{figure}

Fig.\ref{figure5} shows the two normalized interferometric signals in phase quadrature obtained with $^{87}$Rb and $^{85}$Rb for $T=47$ ms. Whatever the acceleration undergone by the atoms, corresponding to a given value of $\alpha$, there is always one of the two interferometers which is working in its linear range (blue-grey circles). Considering a fringe amplitude $A=1$, this linear range corresponds to transition probabilities $P$ such as $\frac{2-\sqrt{2}}{4}<P<\frac{2+\sqrt{2}}{4}$. Selecting from both atom interferometer outputs and at each cycle the measurement corresponding to this linear range gives birth to a fully linear sensor. We can also notice that the correlation between both signals leads to extend the measurement range by a factor 2 compared to a single-species sensor. This is clearly visible on Fig.\ref{figure6} showing the extraction of $\alpha$ for a one atomic species interferometer (red circles) and for the phase shifted dual species interferometer (blue-grey circles).

In order to estimate the gain in sensitivity provided by this configuration compared to a single-species sensor, we performed a numerical simulation. The interferometric phase is uniformly distributed on $\left[0,\pi\right]$ and a random gaussian noise centered on 0 is directly added to the transition probability. The interferometric phases are then extracted thanks to Eq.(\ref{acceleration possible values}).

\begin{table}
\caption{Estimation of the sensitivity gains obtained by coupling the linear ranges of two interferometers in phase quadrature compared to a single species interferometer, as a function of the Signal-to-Noise Ratio (SNR).\label{sensitivity gain}}
\begin{tabular}{cc}
	\hline\hline 
	\rule[1pt]{0pt}{12pt}$~~$SNR$~~$ & $~~$Sensitivity Gain$~~$ \\
	\hline
	\rule[1pt]{0pt}{12pt}10$^2$ & 2.7\\
	10$^3$ & 4.8\\
	10$^4$ & 8.5\\
	\hline\hline
\end{tabular}
\end{table}

The gain in sensitivity is given in TABLE \ref{sensitivity gain}. We notice that the higher the Signal-to-Noise Ratio (SNR), the higher the gain in sensitivity is. Indeed, when the SNR increases, \textit{i.e} when the noise decreases for a given fringe amplitude, the uncertainty on the acceleration estimation decreases much more for points at mid-fringe than for points at the top or at the bottom of the fringes. Typically in our experiment the SNR is $\sim 10^2$, and we expect then a sensitivity gain of $\sim 2.7$. Putting this analysis into perspective, we can assume higher performance atom interferometer exhibiting SNR approaching $10^4$ \cite{Biedermann2009} for which a significant gain in sensitivity, higher than a factor 8, could be reached using this phase shifted dual species interferometer mode.\\

\section{Conclusion}
In conclusion, we have presented and made first analysis of new concepts of cold atom inertial sensor using multi-species atom interferometry. These original configurations allow an improvement of the measurement range or sensitivity compared to standard configurations of a single species atomic accelerometer. For the first configuration, involving interferometers with different interrogation times, the probability noise limits clearly the achievement of more ambitious performances. The second configuration remains very promising to get a fully linear atom interferometer and should be analyzed experimentally more deeply. Note that even if the use of a classical accelerometer is still necessary for the operation of a cold atom accelerometer, this last configuration seems still very interesting to implement. Indeed, this would improve the sensitivity of the global instrument and mitigate the required performance noise of the classical accelerometer. Manipulating simultaneously different atomic species in the same instrument offers new perspectives to be further studied. Using three atomic species such as $^{87}$Rb, $^{85}$Rb and $^{133}$Cs \cite{Diboune2017} would allow for instance to implement simultaneously the two concepts presented in this article, to develop an inertial sensor without deadtimes \cite{Dutta2016} or a simultaneous multiaxis inertial sensor. A strong point of using different atomic species in the same instrument compared to a single species experiment is the ability to increase the number of complementary measurements without disturbance that could arise form spontaneous emission, the lasers manipulating one species being far of resonance relative to the other species.
Ideally, these multi-species concepts should allow to operate a fully atomic sensor without the need to introduce additional vibration isolation platforms or conventional accelerometers. The measurements provided by this kind of instrument could therefore rely only on intrinsic atomic properties and be ultimately limited by quantum projection noise \cite{Leduc2003}.

\bibliography{MultiEspece_v9}

\end{document}